\documentclass[acmsmall,screen]{acmart}
\AtBeginDocument{%
  }

\usepackage{lscape} 
\usepackage{tabularx} 
\usepackage{booktabs} 
\usepackage{svg}
\usepackage{fontawesome}
\usepackage{tabularx}
\usepackage{graphicx} 
\usepackage{multirow}
\usepackage{amssymb}
\usepackage{array}
\usepackage{ragged2e}
\setcopyright{acmlicensed}
\copyrightyear{2025}
\acmYear{2025}
\acmDOI{XXXXXXX.XXXXXXX}

\acmJournal{JACM}
\acmVolume{37}
\acmNumber{4}
\acmArticle{111}
\acmMonth{11}

\begin{document}

\title{Degrading Voice: A Comprehensive Overview of Robust Voice Conversion Through Input Manipulation}

\author{Xinning Song}
\email{xinningsong@tongji.edu.cn}
\affiliation{%
  \institution{Tongji University}
  \city{Shanghai}
  \country{China}
}

\author{Zhihua Wei}
\email{zhihua_wei@tongji.edu.cn}
\affiliation{%
  \institution{Tongji University}
  \city{Shanghai}
  \country{China}
}

\author{Rui Wang}
\email{ruiwang88@iflytek.com}
\affiliation{%
  \institution{iFLYTEK Research}
  \city{Shanghai}
  \country{China}}

\author{Haixiao Hu}
\email{chenyx@hfut.edu.cn}
\affiliation{%
  \institution{Binjiang Institute Of Zhejiang University}
  \city{Zhejiang}
  \country{China}}

\author{Yanxiang Chen}
\email{haixiaohu@sanyau.edu.cn}
\affiliation{%
  \institution{Hefei University of Technology}
  \city{Hefei}
  \country{China}}

\author{Meng Han}
\email{mhan@zju.edu.cn}
\authornotemark[1]
\affiliation{%
  \institution{Zhejiang University}
  \city{Zhejiang}
  \country{China}}

\renewcommand{\shortauthors}{Xinning Song et al.}

\begin{abstract}
  Identity, accent, style, and emotions are essential components of human speech. 
     Voice conversion (VC) techniques process the speech signals of two input speakers and other modalities of auxiliary information such as prompts and emotion tags. It changes para-linguistic features from one to another, while maintaining linguistic contents. 
     Recently, VC models have made rapid advancements in both generation quality and personalization capabilities. These developments have attracted considerable attention for diverse applications, including privacy preservation, voice-print reproduction for the deceased, and dysarthric speech recovery.
     However, these models only learn non-robust features due to the clean training data. Subsequently, it results in unsatisfactory performances when dealing with degraded input speech in real-world scenarios, including additional noise, reverberation, adversarial attacks, or even minor perturbation.
     Hence, it demands robust deployments, especially in real-world settings. Although latest researches attempt to find potential attacks and countermeasures for VC systems,  there remains a significant gap in the comprehensive understanding of how robust the VC model is under input manipulation. There also raises many questions: for instance, to what extent do different forms of input degradation attacks alter the expected output of VC models?  From what perspectives do current defense methods address these attacks, and how can they be categorized based on their defensive state?  Is there potential for optimizing these attack and defense strategies? To answer these questions, we classify existing attack and defense methods from the perspective of input manipulation and evaluate the impact of degraded input speech across four dimensions, including intelligibility, naturalness, timbre similarity, and subjective perception. Finally, we outline open issues and future directions.
\end{abstract}



\maketitle

\section{Introduction}
\label{Introduction}
High-fidelity and personalized audio generation has always been a hot topic in audio domain. Speech synthesis, a task that extracts representational information from various input signals (e.g., voice, language, emotion, songs) and presents them in the form of speech, has attracted widespread attention in society. 
Particularly, voice conversion (VC) is a style-transfer technique that endeavors to transform a source dialect into an expression that resonates with the melodic tones of the target speaker, while retaining the linguistic essence of the source speaker\cite{TodaVC2007}. In other words, VC models modify para-linguistic features such as pitch, timbre and style from source speaker, while preserving speaker-independent information like content.

Typically, a great many VC works can be categorized into three types based on different type of task, that is speaker VC, emotional VC, singing VC. As variants of the basic speaker VC task, emotional VC and singing VC pose greater challenges. Emotional VC focuses on transferring the emotional state provided by the target speaker while preserving other information. Singing VC, on the other hand, places more emphasis on modeling basic components such as pitch, energy, and the singer style. Therefore, both tasks require higher capabilities of feature disentanglement. Specifically,  speaker VC has been extended to miscellaneous tasks such as low-resource VC, dialect conversion, whisper-to-speech, and sub-modules of speech translation.  These rich tasks have captured significant interest in the community, with diverse techniques laying the groundwork for multifaceted applications of voice conversion, including audio editing \cite{wang_audit_2023}, dysarthric speech recovery \cite{wang_duta-vc_2023}, privacy preserving \cite{deng_v-cloak_nodate}, and data augmentation \cite{singh_iteratively_2023}.  


However, these significant VC contributions predominantly relies on clean speech data and often results in unsatisfactory and unexpected performance in real-world scenarios due to the vulnerability of neural networks under additional noise, reverberation, adversarial attacks, or even small mounts of perturbations. 
Recently, adversarial samples generated by $L_\infty $ norm-based constraint \cite{huang_defending_2021,wang2023vsmask}, FGSM \cite{chen2024adversarial}, GAN framework \cite{dong2024active}, frequency band masks \cite{liu2023protecting}, frequency inverse sound pressure levels \cite{yu2023antifake} and psychoacoustics model \cite{li_voice_2023} have proved the possibility of attack on VC models. The most successful adversarial noises can even secretly change the content and speaker identity \cite{chen2024proactive} with transferability on unseen VC models or high efficiency in real-time settings. Besides, additive noises such as white noise, street noise and convolutional noises simulated by pyroomacoustics toolkits serve to introduce distortions to the converted speech data. It leads to a series of bias such as misrepresentation of linguistic content and the obfuscation of affective change, which can potentially influence the effectiveness of individual characteristics manifestation or patient pathological correction. Endeavors such as high-frequency noise elimination \cite{Miaonoise_robust2020}, noise-invariant representation learning \cite{du2022noise,xue22_interspeech}, and cascaded pre-trained model strategies \cite{choi_reverberation-controllable_nodate,choi_evaluation_2022} have been gradually proposed to address these hidden threats. 
These strategies can be further classified into proactive and passive defenses depending on whether they preprocess the degraded input audio. Proactive defenses improve the adaptability of VC model to unknown data by learning robust feature distributions during training, whereas passive defenses ensure that the audio is cleaned before being fed into VC model through various speech enhancement techniques. Despite of these effective ideas, this area still lacks of a definitive framework and comprehensive research. Consequently, it is of great value to categorize the existing works and explore potential strategies to obtain robust VC in real-world settings.

Historically, voice conversion (VC) research has centered on methodology evolution \cite{sisman2020overview}, architecture selection \cite{bargum2024reimagining}, and emerging generative techniques like GANs \cite{dhar2025generative}. Despite recent surveys covering deepfake detection \cite{kim2024comprehensive} and voice cloning terminology \cite{azzuni2025voice}, a dedicated analysis of VC system robustness remains absent. This contrasts sharply with related fields like NLP \cite{wang_measure_2022}, ASR \cite{dua_noise_2023}, and speaker verification \cite{wu2023defender}, where robustness reviews are abundant. As explicitly compared in Table \ref{tab:SurveyCompare}, while other audio domains benefit from systematic security analyses, there remains a distinct scarcity of comprehensive surveys dedicated to the robustness of voice conversion systems. Unlike prior limited discussions \cite{huang_how_2021}, this survey systematically explores three types of speech manipulation techniques, emphasizing adversarial attack strategies and defense characteristics. Our main contributions are summarized as follows.

\begin{itemize}
    \item To the best of our knowledge, this is the first comprehensive survey focusing on the robustness of Voice Conversion models against input data manipulation, thereby directing community attention to this critical yet under-explored domain.
    \item We propose a novel taxonomy of VC vulnerabilities based on input manipulation techniques. Furthermore, we establish a unified evaluation framework that integrates multidimensional metrics—ranging from intelligibility and timbre similarity to subjective perception—to standardize robustness assessment.
    \item We outline potential pathways for robust VC models, including stronger attack strategies that balance imperceptibility, success rate, and transferability, as well as the integration of large-scale speech models with proactive and passive defense strategies. This survey is hoped to provide guidance for the development of secure VC architectures.
\end{itemize}

\begin{table}[!htbp]
    \centering
    \caption{Comparison between this paper and other review articles concerning robustness of different models}
    \label{tab:SurveyCompare}
    
    \resizebox{\textwidth}{!}{
        \begin{tabular}{l c c c l p{8cm}} 
            \toprule
            Authors & Year & Audio & Robustness & Domain & Focus \\
            \midrule
           
            Wang et al. \cite{wang_measure_2022} & 2022 & \faTimes & \faCheck & NLP & A study of NLP robustness definitions, robustness failure identification, and the ways to improve NLP robustness. \\

            Goyal et al. \cite{goyal_survey_2023} & 2023 & \faTimes & \faCheck & NLP & Fulfill the vacuum and classify various defense methods in developing robust and safe NLP tasks \\
            \midrule

            Dua et al. \cite{dua_noise_2023} & 2023 & \faCheck & \faCheck & ASR & Emphasize the importance of noise robust study on ASR and provide an overview of noise resistant techniques, performance metrics, and speech corpus\\
            
            Khan et al. \cite{khan_battling_2023} & 2023 & \faCheck & \faCheck & Voice Spoofing & A study to collate and classify the existing voice spoofing attacks and audio spoofing detection and countermeasures based on traditional and deep learning methods\\

            Li et al. \cite{li_security_2023} & 2023 & \faCheck & \faCheck & Voice Assistant & Outline the security and privacy issues, along with defense methods of ASR and speaker-independent based voice assistant applications \\
            
            Wu et al. \cite{wu2023defender} & 2023 & \faCheck & \faCheck & ASV & A review summarizing current defense strategies against deepfake and adversarial attacks to enhance the robustness of ASV tasks\\

            Lam et al.\cite{kim2024comprehensive} & 2024 & \faCheck & \faCheck & Audio Deepfake & A thorough analysis on current challenges, public datasets, and deep learning techniques in voice conversion, proposing a highly competitive model for Deepfake voice detection task\\

            Zhang et al.\cite{zhang2025audio} & 2025 & \faCheck & \faCheck & Audio Deepfake & A comprehensive survey of advanced deepfake detection and fundamental generation techniques.\\

            \midrule
            
            Sisman et al. \cite{sisman2020overview} & 2020 & \faCheck & \faTimes & Voice Conversion & \textbf{First} review paper traces VC's evolution from statistical methods to deep learning, analyzes key techniques, evaluation methods, reports VCC challenge series, and provides valuable resources for researchers and engineers. \\
    
            Walczyna et al. \cite{walczyna2023overview} & 2023 & \faCheck & \faTimes & Voice Conversion & A review deconstructs Deep Learning-Based Voice Conversion into four key components and presents the latest advancements in each area. \\

            Bargum et al. \cite{bargum2024reimagining} & 2024 & \faCheck & \faTimes & Voice Conversion & A scoping review visualizes research distribution, compares different VC method structures, techniques, and neural sub-blocks, and analyzes reasons behind method selection. \\

            Hussam et al.\cite{azzuni2025voice} & 2025 & \faCheck & \faTimes & Voice Cloning & Explore standardized terminology for voice cloning, and investigate its various variants, as well as related datasets\\

            Dhar et al.\cite{dhar2025generative} & 2025 & \faCheck & \faTimes & Voice Conversion & A comprehensive analysis of voice conversion paradigms is conducted, focusing on the key technologies, major challenges, and transformative impacts of GANs in this field\\

            \midrule
            Huang et al. \cite{huang_how_2021} & 2021 & \faCheck & \faCheck & Voice Conversion & \textbf{First} paper measuring robustness of three typical VC models \\
            
            This paper & 2025 & \faCheck & \faCheck & Voice Conversion & \textbf{First} survey paper of Voice Conversion robustness from the perspective of input manipulation and evaluate the impact of degraded input speech across four dimensions \\
            \bottomrule
        \end{tabular}
    }
\end{table}

The article is organized as follows. Section \ref{sec:background} provides an introduction to the background knowledge of voice conversion system, covering topics such as low-dimensional representation extraction from decoupled speech features, voice conversion tasks, and paralinguistic parameters tuning.
Section \ref{sec:input_manipulation} presents an overview for classifying attacks in voice conversion based on input manipulation. Section \ref{sec:robustness} illustrates the concept of the robust VC system and introduces current passive and proactive defense methods to attacks. Section \ref{sec:benchmark} introduces common datasets, evaluation frameworks and assessment results. Section \ref{sec:open_issue} provides a summary of various VC challenges and discusses the future prospects of robust VC research. Finally, we conclude this survey in the last section.

\section{Voice Conversion}\label{sec:background}



In this section, we provide an overview of the fundamental knowledge surrounding voice conversion systems. Firstly, we present a comprehensive definition of the voice conversion system. Next, we introduce the categories of VC systems, including speaker VC, emotional VC, singing VC, and miscellaneous tasks. Ultimately, we briefly describe the techniques related to speech attributes tuning.

\subsection{Problem Definition}

\begin{figure*}
    \centering
    \includegraphics[width=\linewidth]{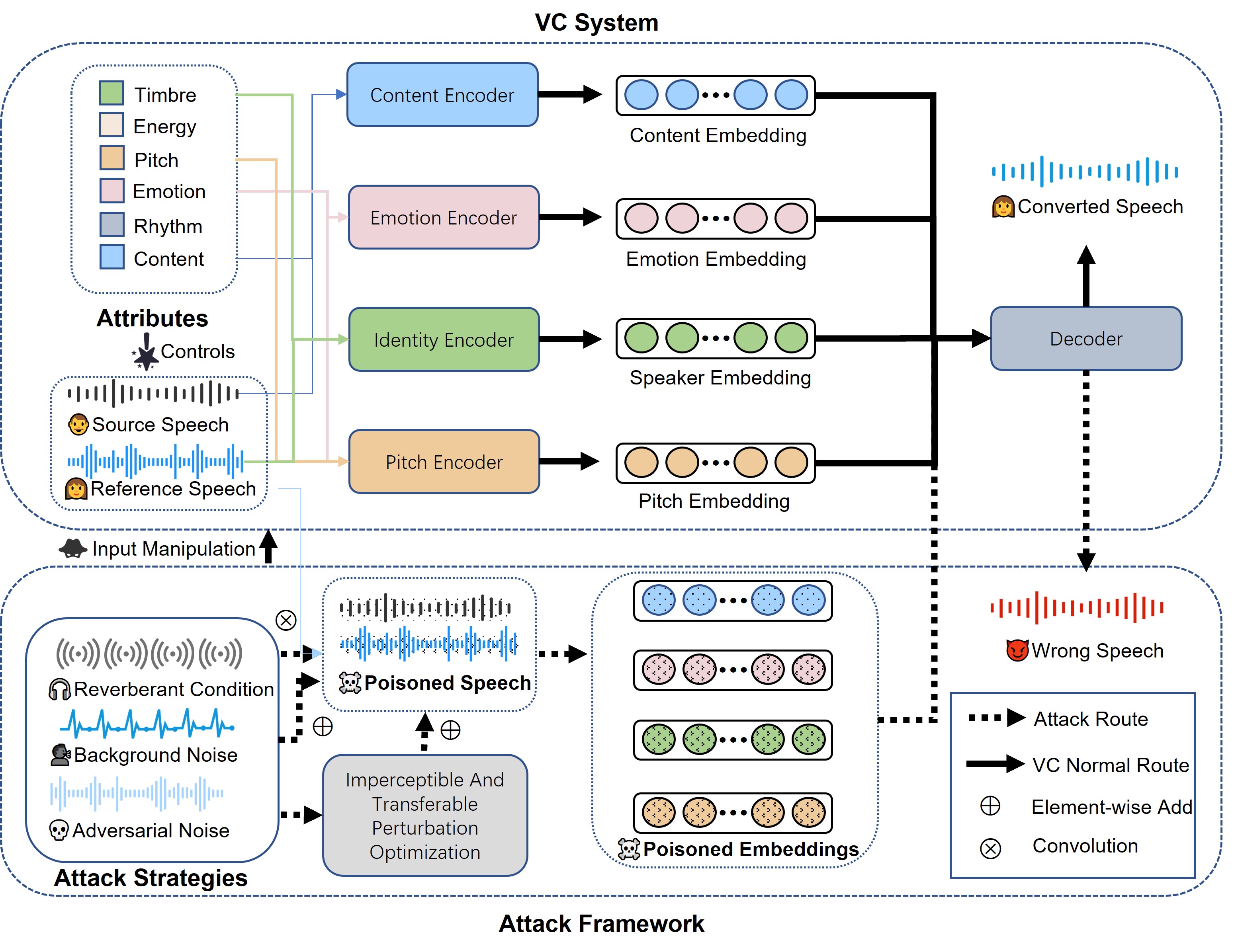}
    \caption{An overview of voice conversion systems and attack framework.}
    \label{fig:VCSSystem}
\end{figure*}

Voice conversion, an instruction technique for para-linguistic information control and linguistic feature refinement, serves as a regression problem \cite{sisman2020overview,zhang2020deepconversion} to find a mapping function between source and target features. Typical voice conversion systems follow the encoder-decoder based architecture depicted in the upper part of Figure \ref{fig:VCSSystem}. Within this paradigm, utterance group and fundamental speech attributes are fed into the VC system,  where they undergo acoustic analysis and reconstruction processes to generate synthesized speech. However, VC models are vulnerable so that the inputs are easily manipulated by various attacks, such as background noise, reverberant conditions, and adversarial noise. These noise signals are convolved or added with the original utterance group, thus imperceptibly contaminating the data. The poisoned speech in the bottom part of Figure \ref{fig:VCSSystem} is subsequently extracted into a variety of contaminated embeddings, recoupled into falsified speech attributes representations, and finally result in generating wrong speech.


Typical VC pipeline integrates various components, including utterance groups, speech attributes, diverse encoders and a decoder. The utterance group comprises source and reference speech, where the source provides content information and the reference determines voiceprint characteristics. Speech attributes, such as time-varying curves\cite{chen_controlvc_2023} or textual prompts, control feature representations extracted from the utterance group. Different VC tasks utilize specialized encoders to decouple inputs into representations like emotion, content, timbre and many other representations. For example, WORLD and YAAPT extract pitch, HuBERT and Paraformer serve as content encoders, and GE2E and ECAPA-TDNN are used for identity extraction. Emotion representation can be explicitly mapped into embeddings or implicitly extracted from audio. To ensure the thorough disentanglement of these representations, techniques, information bottleneck or gradient reversal layer (GRL) are usually introduced. Once these components are effectively decoupled, they are concatenated and fed into the decoder, which synthesizes the target speech characteristics, ultimately contributing to the formation of output audio.

\subsection{Task Categories}

In this subsection, the discussion on task classification is divided into three main categories: speaker VC, emotional VC, and singing VC. Additionally, there are also instances that remain unclassified, such as low-resource settings, dialects, and dysarthric disorders.

Speaker VC is a fundamental task involving the transfer of speaker identity while maintaining linguistic integrity. Although initial research relied on parallel data using GMMs \cite{GMM} and VQ \cite{VQ}, the scarcity of paired data drove the evolution toward non-parallel VC using advanced generative models. Among these, GAN-based methods like StarGAN-VC \cite{kameoka2018stargan} significantly improved naturalness through adversarial training and cycle consistency, though often at the cost of training stability. Alternatively, AutoVC \cite{qian2019autovc} introduced an autoencoder-based framework with a strict information bottleneck, effectively disentangling content and speaker representations to facilitate zero-shot conversion. Recently, the focus has shifted to diffusion models \cite{liu2021diffsvc,popov2021diffusion,popov2023optimal,prabhu2023emoconvdiff}, which formulate VC as a continuous score-based generative process. By utilizing ODE or SDE solvers to reconstruct speech from Gaussian noise conditioned on target features, these models \cite{choi2023dddmvc,guo2023audio} have set new benchmarks in both audio quality and feature disentanglement capability compared to traditional baselines.

Emotional voice conversion (EVC)  \cite{zhou2022emotional} is an advanced task that modifies the emotional state of speech while preserving other voice characteristics. Since emotion is supra-segmental, hierarchical and more complex than neutral talk \cite{Xu_2011}, EVC requires an equal level of attention to both spectral and prosodic elements, unlike speaker VC which places greater emphasis on spectral transformation. Early EVC approaches focus on frame-to-frame mapping of source and target feature to transform the prosody. StarGAN \cite{StarganEVC_2020} is employed to enhance EVC performances by transforming the spectrum envelopes composed of 36 cepstral coefficients. In the face of more challenging scenarios, particularly when target emotional data is absent in both training and testing phases, EVC-USEP \cite{shah_nonparallel_2023} employs two separate encoders and dual-domain source classifiers to learn speaker and emotion embeddings.  It utilizes virtual domain pairing (VDP), fake pair masking (FPM), and simulated annealing to improve results for unseen speaker-emotion pairs.

Singing voice conversion (SVC)\cite{huang2023singingvoiceconversionchallenge} transforms one singer's voice into another’s while preserving the singing content itself. Unlike speaker VC, SVC requires handling a wider range of pitch, energy, and style variations, along with accurate note matching, making it more sensitive to pitch fluctuations and rhythm error. Therefore, the techniques used in speaker VC cannot be directly applied to SVC and it is achieved by accurately simulating pitch and other relevant information. For classic unsupervised SVC model based on autoencoders, speaker confusion techniques help disentangle singer information. PitchNet \cite{Pitchnet_2020} utilizes an adversarial training approach for pitch regression network to learn singer-invariant and pitch-invariant representations. These representations are then fed into the WaveNet decoder to reconstruct the target songs, accurately simulating pitch information to avoid vocal mismatches. UniSinger \cite{hong_unisinger_2023} achieves precise decoupling and control over the semantic, pitch, and timbre information by employing domain adversarial training and mutual information minimization. Besides, information match block is introduced to connect the acoustic model and vocoder, which enables end-to-end training and reduces the distribution gap between the multi-modal inputs.


There are also some miscellaneous atypical VC tasks. In low-resource scenarios, the training set is limited to few samples per speaker, typically not exceeding five samples of 5-10 seconds utterance per speaker \cite{Chen_lowresources2021} or no more than 9 minutes total \cite{baas2020stargan}. Dialect or accent conversion\cite{accentconversion_2020} aims to modify non-native speech to sound native-like, helping second-language learners to listen to their own voices with a native-like accent. In cases of laryngeal disorders or dysarthria, speech intelligibility is reduced due to impairments in pronunciation, phonation,  breathing, resonance, prosody, or their combinations. The whisper-to-speech \cite{perrotin2020glottal} task aims to recreate natural speech from functional whispering produced by patients, restoring a more natural way of speaking. On the other hand, the typical-to-atypical VC task \cite{wang_duta-vc_2023} captures characteristics of speech disorders while preserving typical speaker's identity. It serves as a data augmentation technique to enhance atypical automatic speech recognition capabilities. In noisy background conditions \cite{N2NAPSIPA_Xie,N2N2022_Xie,N2N2023TASLP_Xie}, the combination of linguistic content, speaker identity, and background noise inevitably degrade the similarity and naturalness of transformed speech. In noisy-to-noisy voice conversion task, whether to preserve background noise depends on the specific task and scenario in special environments. For instance, in singing voice conversion task, background sounds carries abundant information, making it essential to preserve them.

\subsection{Speech attributes tuning}
Human speech utterance consists of diverse patterns represented by key attributes such as timbre, energy, pitch, emotion, rhythm, and content. These patterns play a crucial role in concretely representing speech signals. Speech attributes are broadly categorized into speaker-independent linguistic information and speaker-specific para-linguistic components. Semantic content comprises phones that form recognizable formant clusters in the frequency domain, while paralinguistic information includes physical acoustic features, speaker-specific traits, and affective qualities. Physical attributes like pitch and rhythm shape intonation variations and clarity. Timbre serves as a unique vocal fingerprint, and affective attributes,  such as emotion and energy, reflect subjective cognitive intensity.

Speech attribute tuning, including  semantic editing, and attribute-specific control, aims at meeting user-specific requirements in VC tasks, and enhancing the quality and fidelity of synthesized speech. As a fine-grained modification technique, semantic editing allows for precise adjustments through intuitive interfaces or text commands, enabling operations like cutting, copying, pasting, volume adjustment, time stretching, pitch bending, and noise reduction.  However, traditional waveform-based interfaces are less intuitive for non-experts and mismatches between prosody and text often result in audible artifact and unnatural auditory perceptions. To address these problems, VoCo \cite{jin_voco_2017}, a text-based semantic editing system, cascades TTS and VC to match the target timbre. The system allows editors to perform selection, cutting, and pasting operations, with the waveform automatically adjusting accordingly. Recently, novel approaches such as context-aware mask prediction \cite{wang_context-aware_2022} and diffusion-based speech editing \cite{jiang_fluentspeech_2023} are introduced to address two major issues: the unnatural boundaries in the editing region and the inability to synthesize words not present in the transcript. These approaches extend applications to stutter removal \cite{jiang_fluentspeech_2023}, emotion editing \cite{wang2022emotion}, singing voice editing \cite{hong_unisinger_2023} and audio in-painting \cite{wang_audit_2023}.

How to influence different decoupled components by some controllable means is another topic for discussion. In general, there are two types of control in voice conversion: the global and the local \cite{chen_controlvc_2023}. Global control operates  at the utterance level, where the style of the entire sentence is transformed. For instance, input speech can be converted into an exaggerated and expressive statement. In contrast, local control allows frame-level temporal adjustments, modifying detailed speech aspects while preserving natural perception. For these two types, key components primarily include timbre \cite{wu2023transplayer}, pitch \cite{wang_controllable_2022}, emotion intensity \cite{zhou_emotion_2023}, content, prosody \cite{lian_towards_2021}, and rhythm \cite{lee_duration_2022}.


\section{Input Manipulation in Voice Conversion}\label{sec:input_manipulation}
Input manipulation is an intentional act that aims at subverting deep neural networks by distorting input representations to produce unexpected results. In conventional experiments for image generation tasks, interference is typically conducted through various ways, including the addition of noise following specific distributions, adjusting pixel-wise $L_p$ metrics, and training neural networks capable of mapping benign inputs to adversarial samples. However, in  voice conversion tasks, the disturbance affecting generated speech goes beyond adversarial perturbations. The environment reverberation and background noise present in input audio also play a crucial role. Moreover, these three types of input perturbations are more challenging to be added to the pure speech data compared to image data, as the human auditory system is more sensitive to discerning manipulated sound signals than the visual. In this section, we classify the attack strategies in voice conversion tasks into three main categories: adversarial attacks, environment perturbations, and reverberant conditions.

\subsection{Adversarial Attacks}
\label{sub:adversarial attacks}



Adversarial attacks is a classical method of input data manipulation, commonly used in misclassifaction (target attacks) and neural network deception (non-target attacks). It finds minimal perturbations to input data, generating adversarial samples that are imperceptible to humans but capable of altering model outputs. Typically, non-target attacks aims to decrease the similarity between converted utterance and the target in terms of timbre, while target attacks aims at making the converted speech sounds like a third specific speaker. Although adversarial attacks on speech generation models is relatively uncommon, studies like \cite{huang_defending_2021,li_voice_2023} outline the attack process as follows: given a source utterance $x_s$ and a target utterance $x_t$, a content encoder $E_c$, a speaker encoder $E_s$, and a decoder $F$, adversarial attacks introduce subtle perturbations $\delta$ to the original input so that impacts the intended output after passing through the encoders and decoder. As a result, the generated output $F(E_s(x_t+\delta), E_c(x_s))$ becomes out-of-domain compared to the normal generation, consequently rendering the voice conversion task ineffective. As described above, the adversarial sample $\delta$ can be defined as follows:
\begin{equation}
\label{eq:adversarial sample}
        \delta= \epsilon \bigtriangledown_x  J(E_s, E_c,F,x).
\end{equation}
Thus, with the impact of the adversarial sample $\delta$, the perturbed audio $s_{\delta}(t)$ and $t_\delta(t)$ can be represented as follows:
\begin{equation}
\label{eq:adverararial attacks}
\left\{\begin{aligned}
             s_{\delta}(t) = s(t) + \delta \\
             t_{\delta}(t) = t(t) + \delta
\end{aligned}\right.
\end{equation}

Adversarial attacks in voice conversion primarily involve input manipulation in either the frequency or time domains, with the majority of existing research targeting spectral representations to accurately control features such as pitch, rhythm, and timbre. Early frequency-domain methods focused on representation-based optimization; for instance, Huang et al. \cite{huang_defending_2021} systematically investigated end-to-end and feedback strategies to compromise speaker representations in AutoVC \cite{Qian_Zhang_Chang_Yang_Hasegawa-Johnson_2019}, while Chen et al. \cite{chen2024adversarial} applied gradient-based I-FGSM perturbations to the YourTTS encoder \cite{casanova2022yourtts}. However, these optimization-based approaches often suffer from low efficiency and noticeable perturbations. To address these limitations and enhance imperceptibility, subsequent studies have integrated generative models. Dong et al. \cite{dong2024active} innovatively utilized a GAN-based framework to simulate feature extraction for generating high-transferability samples. Similarly, diffusion models have been adopted to preserve audio quality: Wang et al. \cite{wang2025diffattack} proposed DiffAttack using adversarially guided reverse diffusion to maintain timbre, and another study \cite{wang2024diffusion} utilized latent spatial adversarial perturbations (ALP) to further improve robustness and concealment, albeit with potential vulnerability to latent space detection.

Building on these foundations, recent research has extended frequency-domain perturbations to more complex and challenging scenarios, including universal, real-time, and vocal tasks. In the realm of universal attacks, leveraging the foundation established by Cao et al. \cite{cao2021towards}, Ma et al. \cite{ma2025universal} demonstrated a paradigm shift by utilizing adversarial audio prefixes to enable fine-grained, selective control over VC attributes, while Feng et al. \cite{feng2025enkidu} focused on few-shot training to generate cross-model universal perturbations. For real-time applications, Wang et al. \cite{wang2023vsmask} proposed VSMASK to predict effective perturbations for the next time step, masking them in low-frequency bands, though this method lacks adaptive mechanisms. Liu et al. \cite{liu2023protecting} introduced dynamic frequency band masks for online speaker-level transfer. Furthermore, in the specialized domain of voice conversion, Chen et al. \cite{chen2024proactive} (SongBsAb) proposed a scheme balancing lyrics and identity using backtracking maskers and psychoacoustic models, ensuring effectiveness across unknown systems by reducing frame-level interaction.

Conversely, time-domain methods focus on waveform-level perturbations, which can effectively alleviate the impact of upsampling on adversarial performance. These approaches often incorporate psychoacoustic models to balance imperceptibility and adversarial effectiveness. Notably, Voice Guard \cite{li_voice_2023} amplifies attack efficacy using psychoacoustic guidance and introduces metrics like MCD and WER, yet it still faces an inherent trade-off between attack performance and perceptibility. Similarly, Li et al. \cite{li2025cloneshield} optimized perturbations under a multi-objective framework to improve cross-model transfer. Beyond the time domain, addressing black-box constraints remains a critical challenge. To enhance transferability without model gradients, Yu et al. \cite{yu2023antifake} (AntiFake) employed ensemble learning with frequency penalties and the Analytic Hierarchy Process (AHP) to minimize embedding bias, while Gao et al. \cite{gao2025black} tackled the black-box challenge by reducing the search space and refining perturbations with Natural Evaluation Strategies(NES).

In summary, as illustrated in Table \ref{tab:attackMethods}, current adversarial strategies exhibit distinct characteristics: frequency-domain perturbations offer precise feature control, while time-domain methods mitigate upsampling artifacts. Despite significant progress, most existing methods operate primarily at the utterance level under standard white-box or black-box assumptions, often relying on simplistic evaluation metrics. Furthermore, current research faces persistent limitations in transferability to unseen models and robustness against real-world environmental distortions. A critical bottleneck remains in simultaneously achieving high attack efficacy and auditory imperceptibility, as existing techniques struggle to balance these conflicting objectives without compromising high-fidelity generation.

\begin{table*}[!htbp]
\centering
\scriptsize 
\setlength{\tabcolsep}{3pt} 
\renewcommand{\arraystretch}{1.25} 

\caption{\textbf{Summary of Adversarial Attack Methods.}}
\label{tab:attackMethods}

\newcolumntype{S}{>{\hsize=1.35\hsize\RaggedRight\arraybackslash}X} 
\newcolumntype{T}{>{\hsize=0.95\hsize\RaggedRight\arraybackslash}X} 
\newcolumntype{L}{>{\hsize=0.7\hsize\RaggedRight\arraybackslash}X} 
\newcolumntype{G}{>{\RaggedRight\arraybackslash}m{2.1cm}}

\begin{tabularx}{\textwidth}{@{} c c c G S c T L @{}}
    \toprule
    \textbf{Domain} & \textbf{Work} & \textbf{Threat} & \textbf{Granularity} & \textbf{Strategy} & \textbf{Imp.} & \textbf{Transferability} & \textbf{Limitations} \\ 
    \midrule
    
    \multirow{34}{*}{\rotatebox{90}{\textbf{Frequency}}} 
    
    & \cite{chen2024adversarial} & White & \multirow{2}{=}{Offline Utterance} & FGSM and I-FGSM & $\times$ & $\times$ & Simple metrics \\ 
    
    & \cite{dong2024active} & W/B &  & Simulation with GAN-based generation & $\times$ & Substitute model gen. & \\ 
    
    \addlinespace[2pt] 
    \cmidrule(l{2pt}r{2pt}){2-8} 
    \addlinespace[2pt]
    
    & \cite{huang_defending_2021} & W/B & \multirow{2}{=}{Offline Utterance} & Optimization on representation shift & $\times$ & $\times$ & Low efficiency \\ 
    
    & \cite{yu2023antifake} & W/B & & Ensemble learning \& frequency penalties & $\checkmark$ & Test feature similarity & \\ 
    
    \addlinespace[2pt] 
    \cmidrule(l{2pt}r{2pt}){2-8} 
    \addlinespace[2pt]
    
    & \cite{liu2023protecting} & W/B & Offline Utt. / Online Spk. & Masks on frequency band & $\times$ & Cross-model eval. & Complicated optimization \\ 
    
    \addlinespace[2pt] 
    \cmidrule(l{2pt}r{2pt}){2-8} 
    \addlinespace[2pt]
    
    & \cite{wang2023vsmask} & W/B & Online Utterance & Hide perturbations in low frequency band & $\times$ & Cross-model eval. & Single room environment \\ 
    
    \addlinespace[2pt] 
    \cmidrule(l{2pt}r{2pt}){2-8} 
    \addlinespace[2pt]
    
    & \cite{cao2021towards} & White & \multirow{7}{=}{Offline Utterance} & Attacking content via additive noise & $\times$ & Cross-model eval. & Arch. parameters accuracy \\
    
    & \cite{chen2024proactive} & W/B & & Perturbations on lyrics/identity & $\checkmark$ & Test on other encoders & No adaptive attacks \\ 

    & \cite{feng2025enkidu} & White & & Few shot training for frequency perturbations & $\checkmark$ & Cross-model eval. & High-quality box knowledge req. \\ 

    & \cite{gao2025black} & Black & & Reduce search space; refine with NES & $\checkmark$ & Integrate gen. disturbances & Depends on latent model \\
    
    & \cite{ma2025universal} & White & & Add short adv. audio prefix & $\times$ & Cross-model eval. & Specific target attributes \\
    
    & \cite{wang2025diffattack} & W/B & & Adversarially guided reverse diffusion & $\times$ & Cross-model eval. & Speaker constraints \\
    
    & \cite{wang2024diffusion} & W/B & & Stronger perturbations via ALP & $\checkmark$ & $\times$ & Weak spatial robustness \\
    
    \midrule
    
    \multirow{4}{*}{\rotatebox{90}{\textbf{Time}}} 
    & \cite{li_voice_2023} & W/B & \multirow{2}{=}{Offline Utterance} & Optimization on representation shift & $\checkmark$ & $\times$ & Perceptibility trade-off \\
    
    & \cite{li2025cloneshield} & W/B &  & Optimize perturbations under multi-objective & $\checkmark$ & Cross-model eval. & Fails under strong detector \\
    
    \bottomrule
\end{tabularx}

\vspace{0.5ex} 
\begin{minipage}{\linewidth} 
  \footnotesize 
  \raggedright 
  \textbf{Note:} 
  \textbf{W/B}: White-box / Black-box settings; 
  \textbf{Imp.}: Imperceptibility ($\checkmark$: high imperceptibility/inaudible, $\times$: low imperceptibility/audible);
  \textbf{Utt.}: Utterance-level; 
  \textbf{Spk.}: Speaker-level.
\end{minipage}

\end{table*}

\subsection{Environment Perturbations}

In contrast to high-quality and high-sample-rate recordings in laboratories, natural acoustic environments often contain various forms of noise. On the basis of whether the spectral characteristics vary over time, noise can be broadly classified into two categories \cite{zhang_deep_2018}: stationary noise and non-stationary noise. Stationary noise remains relatively constant over time, while non-stationary noise is commonly associated with transient sound events, speaker interference, and music. In practical application scenarios, the source and target speaker signals $s(t)$ and $t(t)$ are often affected by $n(t)$ when propagating through spatial channels. Thus, the modified signals $s_e(t)$ and $t_e(t)$ are defined as:
\begin{equation}
\label{eq:environment perturbations}
\left\{\begin{aligned}
             s_e(t) = s(t) + n(t) \\
             t_e(t) = t(t) + n(t)
\end{aligned}\right.
\end{equation}

In real-world voice conversion tasks, adding additive noise to clean source or target utterances, as illustrated in Equation \ref{eq:environment perturbations}, is a typical way  to simulate environmental noise. Huang et al. \cite{huang_how_2021} introduce pink noise, brownian noise, and indoor noise, with an average level of -30dB to both source and target audio. They evaluate three classic VC models in terms of speaker verification accept rate(SVAR) and character error rate(CER). It is observed that the performance of AdaIN-VC and AUTOVC decreases under brownian noise and pink noise, while DGAN-VC fails to achieve successful conversions. Similarly, Chan et al. \cite{chan2022speech} contaminate clean recordings with "engine", "pink", "white", and "street" noises at SNRs of 5, 10, and 15 dB. Signal-to-Noise Ratio (SNR), a pivotal metric denoting the relationship between signal and noise powers, serves as a fundamental quantitative tool in auditory analysis. SNRs above 0dB indicate a prevalence of signal over noise, whereas SNRs falling below 5dB render signals inaudible.

\subsection{Reverberant Conditions}


Similar to background environmental disturbances, indoor acoustic environments also suffer from the mixture of direct and reflected sound, known as reverberation. When sound propagates indoors, it is absorbed and reflected by obstacles such as ceilings, walls, and floors. The direct sound from the source and the reflected sound reach our ears at different time points, creating a blend of multiple time-point sounds and reflected sounds that shape auditory perception. Assuming direct sound carries maximum energy, while the energy of delayed reflected sounds gradually decreases. Thus, room impulse response (RIR) is introduced to establish temporal and energy relationships between direct sound and subsequent reflections, simulating the reverberation characteristics of a particular space. 

During the spectral analysis of reverberant mixed-noise signals, the reverberation time $T_{60}$, typically around one second is considered along with the duration $L$. When $L$ is much smaller than the analysis window size $T$, the RIR only affects the speech signal within each frame. However, if the reverberation time falls within the range of 200 to 1000 milliseconds, and $L$ is much larger than $T$, it influences the entire speech utterance. In short, given a specific noise signal $r(t)$, and input speech signals $s(t)$ and $t(t)$, we can generate samples with reverberation effects $s_r(t)$ and $t_r(t)$ as follows:

\begin{equation}
\label{eq:reverberant conditions}
\left\{\begin{aligned}
             s_r(t)= s(t)* r(t) \\
             t_r(t)= t(t) * r(t)
\end{aligned}\right.
\end{equation}



In real-world voice conversion tasks, reverberant data is typically generated by mixing clean data with RIRs like Equation \ref{eq:reverberant conditions}. Choi et al. \cite{choi_reverberation-controllable_nodate, choi_evaluation_2022} follow the WHAMR! dataset's reverberation parameter settings and generate reverberant samples by convolving datasets like VCC2018 and PNL100. Their approach results in a significant reduction in the source-to-artifact ratio, scale-invariant signal-to-distortion ratio, and overall speech quality, thereby causing distortion in the converted speech data. Mottini et al. \cite{mottini_voicy_2021} utilize Aachen impulse response (AIR) database and Pyroomacoustics toolkit to simulate reverberation in real-life household environments with three distinct positions for both the sound source and microphones. Takahashi et al. \cite{takahashi_robust_2022} employ a reverberation plugin to generate 20 different RIRs, and apply them to clean singing speech data. This approach increases phonetic feature distance and pitch mean absolute error between the converted utterance and the target. Morita S et al. \cite{2020Voiceauditoryfeedback} convolve the combined signal of air conducted (AC) and bone conducted (BC) sounds with RIR to generate reverberation, resulting in a sharp decrease in the high-frequency components of the converted self-perceived own (SPO) voice. Results show that time delay, AC/BC ratio, and reverberation time can be used to estimate the transfer characteristics between recorded AC and SPO sounds in auditory feedback.


In summary, the input manipulation of source and target speech utterances $s_u(t)$ and $t_u(t)$ can be unified as a combination of adversarial attacks, environment perturbations, and reverberant conditions:

\begin{equation}
\label{eq:unify attacks}
\left\{\begin{aligned}
             s_u(t) = s(t) * r(t) + n(t) + a(\delta,t) \\
             t_u(t) = t(t) * r(t) + n(t) + a(\delta,t)
\end{aligned}\right.
\end{equation}
where $s(t)$ and $t(t)$ symbolize the pristine input speech, $n(t)$ and $r(t)$ refer to two distinct manifestations of natural distribution shift, $a(\delta,t)$ signifies the adversarial samples $\delta$ either in isolation or in conjunction with reverberation $r(t)$.

\section{Robustness in Voice Conversion}\label{sec:robustness}


Adversarial inputs generated by attackers are often concealed yet damaging, causing distortion in voice conversion outputs. Meanwhile, clean speech and mixed-noise speech exhibit highly nonlinear correlations in the time, spectral, power spectrum, mel-spectrum, log-mel-spectrum, or cepstral domains. This inherent nonlinearity makes noise cancellation challenging. Therefore, it is crucial to review existing research and explore robust voice conversion in noisy environments, whether naturally produced or artificially introduced. This section  categorizes defense methods against adversarial attacks and natural distribution shifts into proactive and passive approaches, based on whether they enhance the model's intrinsic robustness to noise or apply denoising techniques prior to the input. Building on this framework, we highlight their connections and common themes. 

\subsection{Robustness in Adversarial Attack}

As discussed in Section \ref{sub:adversarial attacks}, adversarial attacks in VC primarily manipulate input representations to induce target deviations. While defense concepts such as purification have been explored in ASV \cite{wu2023defender}, the generative nature of VC necessitates tailored strategies. Current defenses can be categorized into three paradigms based on their intervention mechanisms: passive purification, proactive robust training, and detection/watermarking.Passive Defense (Purification): Addressing frequency-domain attacks \cite{huang_defending_2021}, Huang et al. \cite{huang_toward_2022} proposed a passive defense utilizing pre-trained Speech Enhancement (SE) modules (e.g., DEMUCS \cite{Défossez_Synnaeve_Adi_2020}, MetricGAN+ \cite{Fu_Yu_Hsieh_Plantinga_Ravanelli_Lu_Tsao_2021}) to filter out adversarial noise $\alpha_{\delta,t}$ before conversion. Recently, more advanced Diffusion-Based Purification methods \cite{2023Diffusion, wang2024defend, fan2025antifake} have been introduced to reconstruct clean audio via controlled noise addition. However, this category serves largely as a remedial, external filtering step, often struggling to balance noise removal with the preservation of high-frequency speech details.

\textbf{Passive Defense And Diffusion Purification:} Addressing frequency-domain attacks \cite{huang_defending_2021}, Huang et al. \cite{huang_toward_2022} proposed a passive defense utilizing pre-trained Speech Enhancement (SE) modules (e.g., DEMUCS \cite{Défossez_Synnaeve_Adi_2020}, MetricGAN+ \cite{Fu_Yu_Hsieh_Plantinga_Ravanelli_Lu_Tsao_2021}) to filter out adversarial noise $\alpha_{\delta,t}$ before conversion. Recently, more advanced Diffusion-Based Purification methods \cite{2023Diffusion, wang2024defend, fan2025antifake} have been introduced to reconstruct clean audio via controlled noise addition. However, this category serves largely as a remedial, external filtering step, often struggling to balance noise removal with the preservation of high-frequency speech details.

\textbf{Proactive Defense And Performance Alignment}: In contrast, proactive approaches integrate defense directly into the model's learning process. Huang et al. \cite{huang_toward_2022} treated adversarial perturbations as data augmentation, employing joint denoising and adversarial training to minimize reconstruction loss. This paradigm has been extended by recent alignment strategies, such as adversarial training \cite{2024Adversarial} and reinforcement learning (RL) based on Supervised Fine-Tuning (SFT) \cite{2025Advancing}, to explicitly align cross-domain perturbations. This internal adaptation makes the model intrinsically robust to input manipulations.

\textbf{Watermarking and Detection}: Given that VC falls under the umbrella of Deepfake technology, establishing the provenance and authenticity of converted speech is critical. Adversaries often exploit this vulnerability by generating "spoofed" samples designed to bypass detectors or strip distinguishing watermarks. To counter such threats, recent research focuses on robust watermarking techniques \cite{2023WavMark, zong2025audiomarknet} and cross-attention based detection \cite{2025XAttnMark} that embed resilient, imperceptible codes. While emerging classifiers \cite{rabhi2024audio} have improved detection accuracy, their robustness is still limited when facing cross-lingual scenarios or sophisticated adaptive perturbations.

\subsection{Robustness under Natural Distribution Shift}

\begin{table*}[!htbp]
    \centering
    \scriptsize 
    \renewcommand{\arraystretch}{1.35} 
    \setlength{\tabcolsep}{3pt} 

    \caption{\textbf{Summary of Defense Methods Against Various Attack Types.}}
    \label{tab:defenseMethods}

    
    \newcolumntype{F}{>{\hsize=1.05\hsize\RaggedRight\arraybackslash}X} 
    \newcolumntype{L}{>{\hsize=0.95\hsize\RaggedRight\arraybackslash}X}
    \newcolumntype{D}{>{\RaggedRight\arraybackslash}m{2.5cm}}
    \newcolumntype{A}{>{\RaggedRight\arraybackslash}m{3.0cm}} 
    \newcolumntype{W}{>{\centering\arraybackslash}m{0.6cm}}  

    \begin{tabularx}{\textwidth}{@{} D W A F L @{}}
        \toprule
        \textbf{Defense Method} & \textbf{Work} & \textbf{Attack Type} & \textbf{Focus} & \textbf{Limitations} \\
        \midrule
        
        Spectral Subtraction & \cite{Miaonoise_robust2020} & Environment Perturbations & Uses low-pass filtering to remove high-frequency noise and filters MCEPs. & Unable to perform zero-shot conversion; weak baseline performance. \\
        \cmidrule{1-5} 

        \multirow{4}{=}{Adversarial Training} 
        & \cite{du2022noise} & Environment Perturbations & Introduces GRL and domain classifiers for noise invariance. & High computational cost; only resists identified attacks. \\
        & \cite{2024Adversarial} & Speaker Embeddings & Incorporates adversarial examples directly into training. & Limited generalization ability towards unknown attacks. \\ 
        \cmidrule{1-5}

        \multirow{5}{=}{Denoising Training} 
        & \cite{xue22_interspeech} & Environment Perturbations & Encoder-decoder utilizing FLIM for noise control. & Insufficient decoupling of features (identity, pitch, content). \\
        & \cite{mottini_voicy_2021} & Reverberant Conditions & Encoder-decoder adding extra phonetic and acoustic-ASR modules. & Prosodic leakage and information loss during phoneme transcription. \\
        \cmidrule{1-5}

        \multirow{5}{=}{Cascade SE Model} 
        & \cite{choi_reverberation-controllable_nodate} & Reverberant Conditions & T60 estimator and VAE-based separation framework. & Residual reverberation remains and affects VC performance. \\
        & \cite{choi_evaluation_2022} & Noisy \& Reverberant & Three-stage framework using two independent pre-trained SE modules. & Experiments limited to single-channel and 8kHz sample rate. \\
        \cmidrule{1-5}

        \multirow{7}{=}{Diffusion-based Purification} 
        & \cite{2023Diffusion} & Adversarial Attack & Reconstructs clean audio via reverse diffusion with controllable noise. & Effectiveness depends heavily on training levels and attack settings. \\
        & \cite{wang2024defend} & Adversarial Attack & Active perturbation purification based on diffusion models. & High computational overhead; relies on parameter accuracy. \\
        & \cite{fan2025antifake} & Adversarial Attack & Initial diffusion purification with progressive phoneme-guided refinement. & Security protection is incomplete with potential privacy risks. \\
        \cmidrule{1-5}

        \multirow{7}{=}{Robust Watermarking} 
        & \cite{2023WavMark} & Environmental \& Signal & Embeds invisible watermarks to identify provenance. & Vulnerable to strong audio pre-processing. \\
        & \cite{2025XAttnMark} & Transform \& Editing & Cross-attention architecture for reliable detection. & Accuracy under extreme adaptive attacks is pending. \\ 
        & \cite{zong2025audiomarknet} & Adaptive Attack & Embeds semantic watermarks in original speech. & Robustness against physical playback is untested. \\
        \cmidrule{1-5}

        Preference Alignment & \cite{2025Advancing} & Cross-domain Perturbation & Introduces SFT and reinforcement learning (RL) for alignment. & Does not cover specialized terminology or rare language pairs. \\
        \cmidrule{1-5}

        Classifier Defense & \cite{rabhi2024audio} & Adversarial Attack & Advanced audio deepfake classifiers to detect adversarial inputs. & Robustness against all attack types has not been fully explored. \\
        \cmidrule{1-5}

        Hybrid Defense & \cite{huang_toward_2022} & Unified Degradation & Comprehensive study using pre-trained SE models and denoising training. & Denoising training lowers naturalness; SE models cause distortion. \\

        \bottomrule
    \end{tabularx}
\end{table*}

As shown in Equation \ref{eq:unify attacks}, natural distribution shift is determined by environment noise $n(t)$ and reverberation $r(t)$ if not concern about adversarial noise $\delta$. Therefore, an ideal robust VC model should eliminate $r(t)$ and $n(t)$ to recover clean input data $s(t)$, $t(t)$. Built upon this principle, existing research primarily revolves around two approaches: learning noise-robust representations and utilizing pre-trained denoising or dereverberant models to preprocess input signals with two types of disturbances. For the first method, Mottini et al. \cite{mottini_voicy_2021} introduces an additional phonetic encoder to encode sentence representations into text, along with incorporating speech embeddings predicted by an ASR module to enhance clarity. Du et al. \cite{du2022noise}  introduces a gradient reversal layer (GRL) and domain classifier modules to reduce the gap in representations between noisy and clean data domains. Through domain adversarial training, the learned content and speaker representations from noisy and clean speech are respectively enforced to be noise-invariant.

For the second approach, Xie et al. \cite{N2NAPSIPA_Xie,N2N2022_Xie} propose  a noisy-to-noisy (N2N) voice conversion framework to adapt three different noise conditions: speaker-independent, semi-speaker-dependent, and speaker-dependent scenarios. This framework preserves background additive noise during conversion. Initially, a pre-trained speech enhancement (SE) model DCCRN \cite{Hu2020DCCRNDC} separates the noisy speech into speech and noise components. Additionally, to prevent information leakage of speaker identities into noise condition, they introduce three data augmentation strategies: Data-Aug, Noise-Aug I, and Noise-Aug II \cite{N2N2023TASLP_Xie}. Data-Aug expands noise diversity by mixing original training data with noise at varying signal-to-noise ratios (SNRs).  Noise-Aug I only utilizes the augmented enhanced noisy data to compute the loss, while keeping original training data processed by SE module to reduce the impact on the quality of the training set. Noise-Aug II duplicates and combines the denoised speech processed by SE module with augmented noise segments to form enhanced noisy speech, while ensuring the decoupling of speaker-dependent noise. These strategies are proved to be necessary as it would otherwise result in a significant decrease in the naturalness and similarity of speech generated by VQ-VAE \cite{VQVAE_2020} model in the second stage. For more challenging noise-reverberation scenarios, Choi et al. \cite{choi_evaluation_2022} propose a cascaded voice conversion framework that  employs pre-trained denoising and dereverberation models. Results show that incorporating either SE model for data preprocessing can enhance the performance of VC models, with the combined utilization of both models yielding even better results.


\subsection{Connections and A Common Theme}

The robust solutions for real-world VC scenarios discussed previously can be unified into a single framework that provides robust defense strategies against input manipulations caused by adversarial attacks or natural distribution shifts.  It ensures that the altered pure input signals remain unaffected, even when they fall into degraded data. As shown in Table \ref{tab:defenseMethods}, the primary robust VC methodologies include spectral subtraction, adversarial training, denoising training, cascading SE model, and hybrid defense. These methods are categorized into passive and active defense strategies. Passive defense strategies aim to enhance input data quality by reducing noise and other distortions, while active defense strategies aim to make the model more robust against adversarial attacks and natural variations in data distribution.

A classic passive defense approach is speech enhancement, including spectral subtraction \cite{Miaonoise_robust2020} and cascading the pre-trained SE models \cite{choi_reverberation-controllable_nodate,choi_evaluation_2022}. Spectral subtraction, a statistical learning method, involves pre-process and post-process steps to eliminate high-frequency noise and augment speech quality. Cascading SE model, on the other hand, employs sequential enhancement stages to purify noisy speech before passing it to VC modules. Both models are straightforward and intuitive for basic noise reduction, they still present certain limitations and challenges in more complex scenarios. Spectral subtraction is primarily designed to handle natural variations rather than artificial perturbations. The solution struggles to manage scenarios with more complex coupled speech features, such as speaker emotion and pitch. Cascading SE model increases a set of parameters which is unfriendly to real-time scenarios or mobile platforms. Moreover, SOTA SE model may eliminate essential VC features or even cause output distort due to the mismatched evaluation goal between SE and VC tasks.

Proactive defense methods fall into two categories: one involves conducting denoising and adversarial training directly on noisy speech or adversarial samples, while the other focuses on employing domain adversarial training to push the model to learn noise-invariant representations. However, these approaches still face challenges such as feature leakage and residual noise. Overall, further refinement of passive and proactive strategies is essential to balance VC robustness with these limitations.

\section{Datasets and Evaluation}\label{sec:benchmark}

The foundation of a successful VC system lies in the dataset, while the assessment of a VC system's quality requires evaluation metrics. In this chapter, we introduce commonly used clean and environmental noise datasets for VC systems and outline key subjective and objective metrics for assessing their robustness. 

\subsection{Datasets}
Developing a robust VC system requires two types of datasets: clean VC speech datasets and noisy speech datasets. Clean datasets exclude low-quality utterances, while noisy datasets incorporate various disturbances, such as additive and RIR noises, to simulate natural environmental conditions.

Key clean VC speech datasets include VCTK \cite{Veaux2016SUPERSEDEDC}, LibriTTS \cite{Zen2019LibriTTSAC}, VCC2018 \cite{LorenzoTrueba2018TheVC}, and AISHELL series \cite{Bu2017AISHELL1AO,Fu2021AISHELL4AO}.
\begin{itemize}
    \item \textbf{VCTK}: This corpus encompasses audio recordings from 110 English speakers, each delivering approximately 400 sentences from newspapers, the Rainbow Passage, and elicitation paragraphs. These audio are captured at a sample rate of 96kHz using an omnidirectional microphone (DPA 4035) and a small diaphragm condenser microphone with an exceptionally wide bandwidth (Sennheiser MKH 800). 
    \item \textbf{LibriTTS}: A high-quality, multi-speaker English corpus with approximately 585 hours of read English speech at 24kHz. The dataset improves upon the original corpus LibriSpeech by excluding sentences with significant background noise and segmenting audio at natural pause points. Both original and normalized text are included for context extraction.
    \item \textbf{VCC2018}: Datasets developed for the famous competition "Voice Conversion Challenge 2018". It includes parallel utterances for training conversion models in Hub task and non-parallel utterances for performance evaluation in Spoke task.
    \item \textbf{AISHELL series}: Mandarin speech corpora created by Beijing Shell Technology Co., featuring diverse accents. AISHELL1 offers 520 hours of speech from 400 speakers. AISHELL3 provides 85 hours and 88,035 sentences  of high-fidelity recordings, while AISHELL4 includes 120 hours of multi-channel speech from 211 conferences involving 4–8 participants.
\end{itemize} 

Noisy Speech Dataset can primarily be categorized into two major types: pure environmental noise datasets devoid of human speech and datasets directly recorded with human speech interference under noisy conditions. The former comprises datasets such as PNL 100 \cite{PNL100}, DEMAND \cite{DEMAND}, ESC50 \cite{piczak2015dataset}, and Noisex92 \cite{varga1992noisex}.
\begin{itemize}
    \item \textbf{PNL 100}: A diverse collection of daily nonspeech noises, including but not limited to machine noise, bells, alarms, yawns, showers, traffic, and car noise. It can be used to evaluate noise-robust VC systems.
    \item \textbf{DEMAND}: A corpus proposed to address the limited environmental diversity and sparse artificial noise sources provided by previous datasets. This dataset includes 16-channel recordings with microphones placed 5–21.8 cm apart, offering genuine noise recordings from various settings rather than simulated environments.
    \item \textbf{ESC-50}: A curated set of 2000 five-second single-channel noisy audio recordings categorize into 50 semantic classes, covering a wide range of natural, human, and domestic sounds.
    \item \textbf{NOISEX-92}:A database of 16-channel environmental noise recordings, available on two CD-ROMs. It includes a wide range of sounds such as voice babble, factory noise, HF radio channel noise, pink noise, white noise, and various military noises.
\end{itemize}
The latter encompasses datasets like WHAMR! \cite{Maciejewski2019WHAMRNA} and CHiME \cite{barker2017chime}, which is under reverberation conditions and noise disturbances.     
\begin{itemize}
    \item \textbf{WHAMR!}:An extension of the WHAM!, which is designed for noisy and reverberant speech seperation tasks. It utilizes pyroomacoustics to generate room impulse responses and incorporates synthetic reverberation that mimic domestic and classroom environments with varying reverberate levels.
    \item \textbf{CHiME}: A series of datasets and challenges that have evolved to include increasingly complex scenarios.  Early CHiME datasets focus on single-channel, small-vocabulary tasks in controlled environments, while later versions expand to multi-channel, large-vocabulary tasks in more diverse and unpredictable environments. 
\end{itemize}

Additionally, toolkits like the Adversarial Robustness Toolbox (ART) facilitate generating audio adversarial examples. This Python library, developed for machine learning security, supports evaluating, defending, and certifying machine learning models against threats such as evasion, poisoning, and inference attacks. It also provides two examples\cite{qin2019imperceptible,AudioAdversarialExamples} for crafting imperceptible targeted audio adversarial examples for speech recognition systems,

\subsection{Evaluation Metrics}
\label{subsec:evaluationmetrics}
In recent years, the growing interest in VC research has underscored the need for consistent quality validation of generated speech and robust comparisons among VC systems. To address this, a unified set of evaluation metrics is essential. VC evaluation can be broadly categorized into four dimensions shown in Table \ref{tab:MetricsSummary}: intelligibility, naturalness, timbre similarity, and subjective perception. Below, we introduce key metrics for assessing robust voice conversion.

\begin{table*}[!htbp]
    \caption{Summary of Metrics on Robust Voice Conversion}
    \label{tab:MetricsSummary}
    \begin{tabular}{ccccc}
        \toprule
          & \textbf{Intelligibility} & \textbf{Naturalness} & \textbf{Timbre similarity} & \textbf{Subjective Perception} \\
        \midrule 
         $WER$ & \faCheck &  &   &   \\
         $MCD$ &  &\faCheck  &   &   \\
         $F0RMSE$ &  &\faCheck  &   &   \\
         $F0CORR$ &  &\faCheck  &   &   \\
         $SS$ &  &  & \faCheck   &   \\
         $EER_{th}$ &  &  & \faCheck   &   \\
         $EER$ &  &  & \faCheck   &   \\
         $ASR$ &  &  & \faCheck   &   \\
         $UTMOS$ &  &  &    & \faCheck   \\
        \bottomrule
    \end{tabular}
\end{table*}




1$)$ Intelligibility: \textbf{Word error rate(WER) } is a typical method for evaluating speech intelligibility. It measures the comprehensibility of generated speech by calculating the edit distance between the target transcription and the reference sequence. Given the total number of words in original utterance $N$, the edit distance is determined by the minimum number of operations between original utterance and converted utterance transcriptions, including substitutions($S$), deletions($D$), or insertions($I$). Hence, WER can be computed using the following formula:


\begin{equation}
\label{eq:WER}
WER = \frac{I+D+S}{N} \times 100\%
\end{equation}

We note that  a lower WER (Word Error Rate) indicates fewer errors in the machine's recognition of generated speech, such as spelling mistakes or semantic issues, which in turn reflects better audio intelligibility.

2$)$ Naturalness: The perception of voice quality, primarily shaped by factors such as pitch contours and multi-dimensional coefficients, reflects how natural an utterance sounds. Metrics like Pearson Correlation Coefficient (PCC), Root Mean Squared Error (RMSE), and Cepstral Distance(CD) are commonly employed to measure the relative distance between generated audio and reference audio. Furthermore, \textbf{F0RMSE}, \textbf{F0CORR}, and \textbf{Mel-cepstral distortion (MCD)} account for both ptich and acoustic coefficient vectors, which influence vowel dispersion and overall speech quality.

\textbf{Mel-cepstral distortion (MCD)}\cite{kubichek1993mel, kominek2008synthesizer} is a metric aligned with the variation of the ear's critical bandwidth, whose core principle lies in calculating the Euclidean distance the mel-frequency cepstral coefficients (MFCCs) of the converted audio and the reference audio. Given the $i-th$ coefficient at $k$ frame, $m_{ref}^{(k,i)}$(the reference audio) and $m_{conv}^{(k,i)}$ (the converted audio), MCD is computed:

\begin{equation}
\label{eq:MCD}
MCD[dB] = \frac{10\sqrt{2}}{ln 10} \frac{1}{N} \Sigma_{i=0}^{N-1} \sqrt{\Sigma_{k=i} ^{T} {(m_{conv}^{(k,i)}-m_{ref}^{(k,i)})^2}}
\end{equation}
where $T$ represents the length of Mel-cepstral coefficients vectors and $N$ denates frames numbers. A lower MCD value indicates the generated utterance is acoustically closer to the reference. In other words, it sounds more natural to listeners. Specifically, when the lengths of two speech sequences differ, dynamic time warping (DTW) can be applied to align them for accurate comparison.

\textbf{Fundamental frequency metrics F0 RMSE \&  F0 CORR}: Fundamental frequency (F0) features  olay a pivotal role in assessing the naturalness of generated speech, as they effectively capture the variation in voice prosody across pitch periods. To justify the naturalness of F0 contours, dynamic time warping(DTW) is applied to align F0 features extracted from the reference and generated audio. Once aligned, Pearson Correlation Coefficient and Root Mean Squared Error are introduced to measure the distance:

\begin{equation}
\label{eq:F0CORR}
F0CORR = \frac{Cov(F0^{ref},F0^{conv})}{\sigma_{F0^{ref}}\sigma_{F0^{conv}}}
\end{equation}

\begin{equation}
\label{eq:F0RMSE}
F0RMSE[Hz] = \sqrt{\frac{1}{N}\Sigma_{i=1}^{N}(F0_i^{ref})-(F0_i^{conv})^2}
\end{equation}
where $\sigma_{F0^{ref}}$, $\sigma_{F0^{conv}}$ refers to statistics standard deviation of the fundamental frequency sequence $F0^{ref}$ and $F0^{conv}$, while $N$ represents the length of F0 sequence. We note that a value of F0CORR closer to 1 or a smaller F0 RMSE indicates that the generated audio is more similar to the reference, reflecting a higher degree of naturalness in the generated result

3$)$ Timbre Similarity: 
Timbre is a crucial vocal feature reflecting speaker identity. \textbf{Speaker Similarity (SS)}, calculated as the cosine similarity between timbre features extracted by pre-trained speaker recognition models, is a classic method for determining whether the generated identity in converted speech matches the expected one.

\begin{equation}
\label{eq:SS}
SS = \frac{E_s(U^{ref}) \cdot E_s(U^{conv})}{|E_s(U^{ref})| |E_s(U^{conv})|}
\end{equation}
where $E_s(\cdot)$ refers to the pre-trained speaker recognition models like Dvector\footnote{https://github.com/yistLin/dvector}. $U^{ref}$ and $U^{conv}$, respectively, represent the reference and the converted utterance. 

\textbf{Equal error rate (EER)} is a common metric to evaluate the accuracy of speaker recognition models. A lower EER value indicates the model is more suitable for the task. Further, we treat \{adversarial audio, attack audio\} as positive samples and \{adversarial audio, original audio\} as negative samples,  identifying the decision threshold \textbf{$EER_{th}$} at the point where the False Reject Rate(FRR) equals the False Accept Rate(FAR).

\textbf{Attack Success Rate(ASR)} stands for the likelihood that a voice conversion model’s output will be altered by adversarial audio. Based on the similarity value at the balance point $EER_{th}$, if the cosine similarity between the converted output audio and the attack audio exceeds this threshold, they can be considered as having the same speaker identity, indicating a successful attack. By counting the number of successful attacks $N_{th}$, given the total number $N$, the Attack Success Rate (ASR) can be calculated using the following formula:

\begin{equation}
\label{eq:ASR}
ASR = \frac{N_{th}}{N}
\end{equation}

4$)$ Subjective Perception: \textbf{AB preference tests} and \textbf{Mean Opinion Score (MOS)} are widely-used methods for subjective perception assessment, providing insights into user preferences and perceived quality in speech evaluation. AB preference tests are qualitative methods designed to understand why users prefer one option over another. Participants are asked to choose which of two audio samples—converted or reference— better demonstrates a specific attribute. In MOS test, participants rate speech quality on a scale of 1 to 5 under uniform scoring criteria. Generally, a MOS score above 4 is considered high-quality, while a score below 3.6 indicates unsatisfactory quality.

\textbf{UTMOS}\cite{saeki2022utmos} is a MOS prediction system which proposed in VoiceMOS Challenge 2022. In practical VC scenario, this system is used for automatically evaluating the converted speech quality.

Moreover, specific metrics are employed in particular tasks, such as assessing emotional intensity and measuring the similarity between the target and converted emotions in tasks like emotion voice conversion or emotional speech synthesis. Additionally,  for tasks such as singing voice conversion, singing speech synthesis, or singing voice editing, F0 frame error is commonly employed to assess pitch accuracy.


\subsection{Evaluation Experiments}

In this part,  we conduct several experiments to evaluate the impact of three types of input manipulation on voice conversion performance across four evaluation dimensions outlined in Section 
\ref{subsec:evaluationmetrics}. All experiments are performed using a single NVIDIA GTX 4090 GPU.

\begin{figure*}
    \centering
    \includegraphics[width=\linewidth]{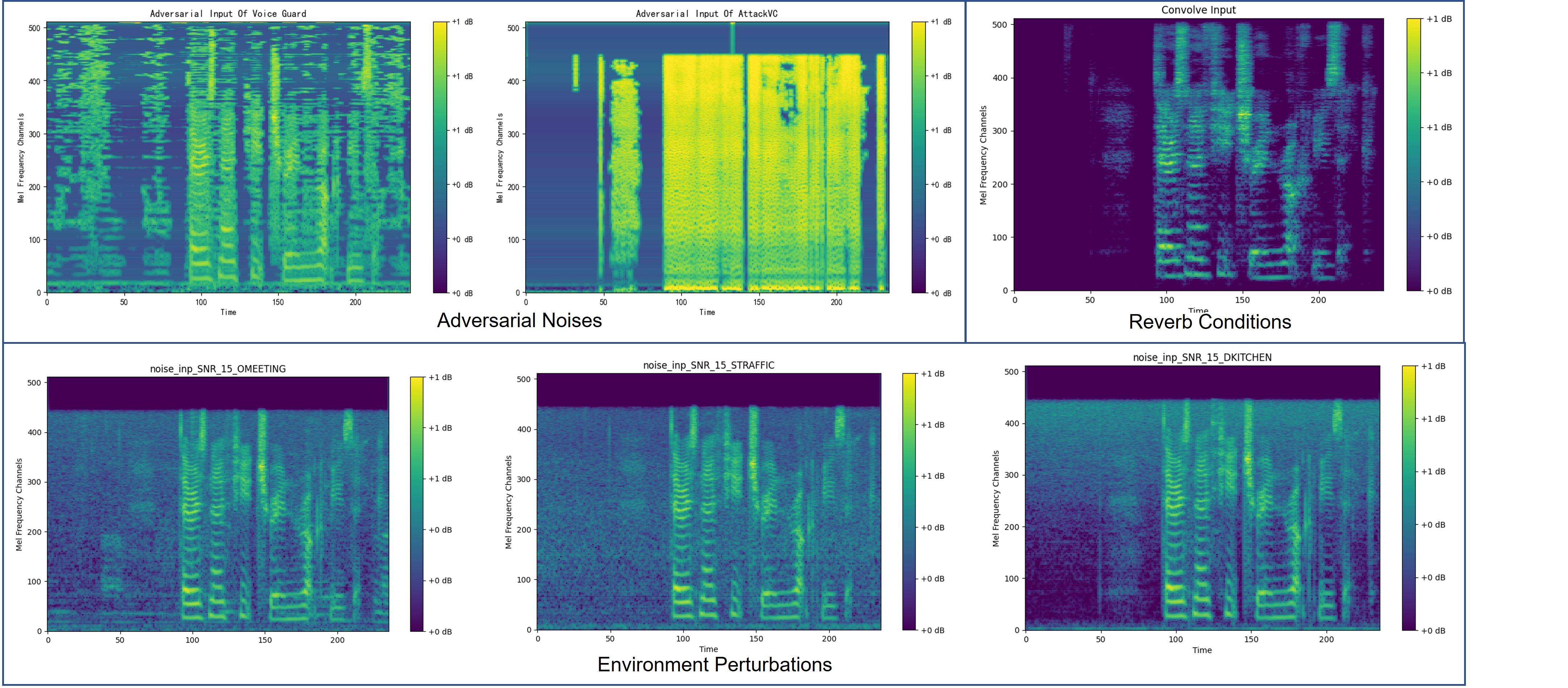}
    \caption{Mel spectrograms of three different degraded noises.}
    \label{fig:DegradedNoise}
\end{figure*}

\textbf{1$)$} Experiment Setup: We adopt AdaIN-VC, a classic zero-shot voice conversion model for this study. For datasets under three degradation conditions discussed in Section \ref{sec:input_manipulation}, we select VCTK\cite{Veaux2016SUPERSEDEDC} as clean sample group, DEMAND as noise sample group, and generate a reverberant sample group using the RIR Generator with fixed room dimensions and positions of the sound source and receiver.

Specifically, for adversarial perturbation scenario, we randomly selected 250 audio pairs from the VCTK dataset, each consisting of an attack utterance, an original utterance, and a content utterance. Our experiment followed three key criteria:  (1) each audio file must avoid extended pauses at the beginning and end, with a speaking rate above 0.6; (2) speakers for each pair are selected so that the attack and original audios came from speakers of different genders (e.g. if the attack utterance is from a female speaker, the original utterance will be from a male speaker); and (3) each pair includes one long audio (6–8 seconds), one medium-length audio (4–6 seconds), and one short audio (<4 seconds), all from three different speakers.  The adversarial attack assessment is carried out on AttackVC\cite{huang_defending_2021} and VoiceGuard\cite{li_voice_2023}.

For environmental noise scenarios, we select three different noise types from DEMAND dataset: indoor noise environment (DKITCHEN), background speaker noise environment (OMEETING), and outdoor environment (STRAFFIC). Taking DKITCHEN as an example, the noise mixing process follow these steps: (1) randomly selecting an $SNR_{db}$ value from [5,10,15]; (2) performing voice activity detection (VAD) on a 5-minute noise audio segment (randomly chosen from $ch_01$ to $ch_16$) to identify active frames; and (3) using a fixed random seed of 42 to ensure the processed noise audio frames is of the same length of the original utterance for additive mixing, resulting in 250 mixed audio samples. 

Finally, the reverberation scenario is configured with a room size of $5\times4\times6$, a default sound source position of $[2, 3.5, 2]$, and a default three-channel receiver positioned at $ [[2, 1.5, 1], [2, 1.5, 2], [2, 1.5, 3]]$.

\textbf{2$)$ Experiment Results:} We begin by using mel-spectrograms to analyze the differences and connections among Adversarial Noises, Reverb Conditions, and Environmental Perturbations, as shown in Figure \ref{fig:DegradedNoise}. These noise  types are generated according to the configurations described above. Notably, adversarial noise heavily overlays the original audio, while environmental noise manifests as small, dense patches on the clean audio. In the reverberation scenario, the upper formant information is visibly degraded. Specifically, for adversarial noises with greater imperceptibility, such as in the top-left image in Figure \ref{fig:DegradedNoise}, the noise overlay is lighter and less intrusive.

\begin{table*}[!htbp]
\centering
\caption{Evaluation On Two Classic Adversarial Methods}
\label{tab:Adversarial}
\resizebox{\textwidth}{!}{ 
\begin{tabular}{|c|c|ccc|ccc|c|}
\hline
\multirow{2}{*}{\textbf{Attack Scenario}} & {\textbf{Intelligibility}} & \multicolumn{3}{c|}{\textbf{Naturalness}} & \multicolumn{3}{c|}{\textbf{Timbre Similarity}} & \textbf{Subjective Perception} \\
\cline{2-9}
& \textbf{WER} & \textbf{MCD} & \textbf{F0RMSE} & \textbf{F0CORR} & \textbf{EER} & \textbf{$EER_{th}$} & \textbf{ASR} & \textbf{UTMOS} \\
\hline
Frequency Domain AttackVC\cite{huang_defending_2021}& 4.7\%$\rightarrow$4.9\% & 5.94 $\rightarrow$ 6.76 & 26.89 $\rightarrow$ \textbf{60.72} & 0.63 $\rightarrow$ \textbf{0.45} & 1.80\% & 0.9200 & \textbf{95.2\%} & 0.73 $\rightarrow$ 0.68 \\
\hline
Time Domain VoiceGuard\cite{li_voice_2023} & 4.7\%$\rightarrow$4.9\% & 5.94 $\rightarrow$ \textbf{6.33} & 26.89 $\rightarrow$ \textbf{46.17} & 0.63 $\rightarrow$ 0.58 & 40.28\% & 0.9250 & 51.4\% & 0.74 $\rightarrow$ 0.70 \\
\hline
\end{tabular}
} 
\end{table*}

\begin{figure*}
    \centering
    \includegraphics[width=\linewidth]{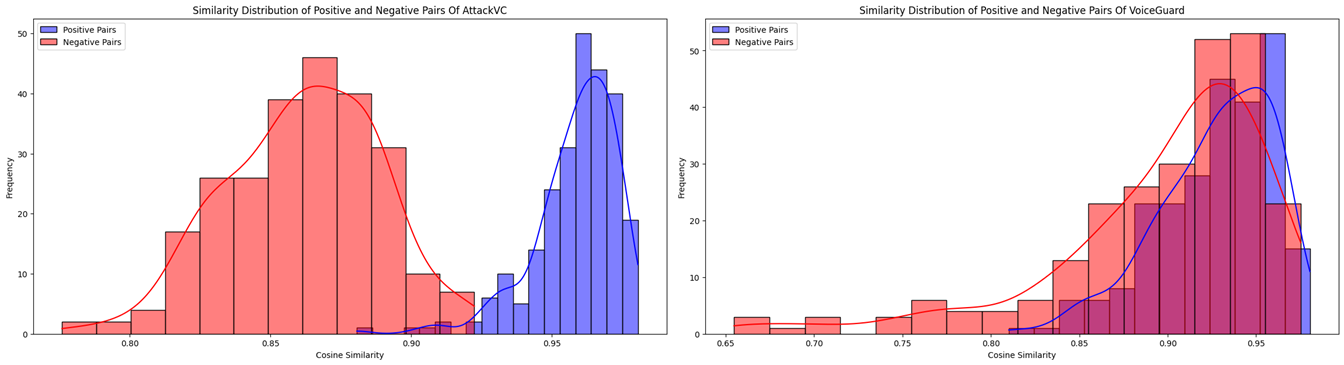}
    \caption{Equal Error Rate Distribution Of Two Adversarial Attack Methods.}
    \label{fig:EERDistribution}
\end{figure*}

\textbf{Adversarial Noise}: We only present evaluations on AttackVC\cite{huang_defending_2021} and VoiceGuard\cite{li_voice_2023} in Table \ref{tab:Adversarial}, as they represent the time-domain and frequency-domain adversarial attack methods for voice conversion, respectively. Basically, Word Error Rate (WER) of the converted and content utterance before and after applying two adversarial attack methods exhibits a slight increase of only 0.2\%. This indicates current adversarial perturbation techniques are able to maintain imperceptibility. In terms of naturalness, the Mel Cepstral Distortion (MCD) shows minimal change before and after applying the two attack methods, which means minimal quality distortion. However, pitch-related metrics exhibit more noticeable variations. For instance, the F0RMSE value increases significantly from 26.89 to 60.72 with the AttackVC algorithm, accompanied by a drop in F0CORR of nearly 0.2. VoiceGuard, with better adversarial imperceptibility, exhibits relatively smaller changes but still notable. The timbre similarity experiment results, shown in Figure \ref{fig:EERDistribution}, reveal an Equal Error Rate (EER) threshold of approximately 0.92. At this threshold, only 51.4\% of perturbation-generated converted utterance exhibit a cosine similarity above the threshold with the attack utterance. Results show that VoiceGuard achieves a lower attack success rate, demonstrating a trade-off between imperceptibility of adversarial attack and its success rate. Finally, UTMOS decreases by only 0.05 in subjective perception evaluations for both adversarial attack models. This may be due to AdaIN-VC, as an early model, producing relatively low-quality converted utterances, so changes to speaker identity have minimal perceptual impact.

\begin{table*}[!htbp]
\centering
\caption{Evaluation On Environment Perturbations}
\label{tab:Environment Perturbations Evaluation}
\resizebox{\textwidth}{!}{ 
\begin{tabular}{|c|c|ccc|c|c|} 
\hline
\multirow{2}{*}{\textbf{Environment Scenario}} & \textbf{Intelligibility} & \multicolumn{3}{c|}{\textbf{Naturalness}} & \textbf{Timbre Similarity} & \textbf{Subjective Perception} \\
\cline{2-7} 
& \textbf{WER} & \textbf{MCD} & \textbf{F0RMSE} & \textbf{F0CORR} & \textbf{SS} & \textbf{UTMOS} \\
\hline
OMEETING & 4.7\%$\rightarrow$ 
\textbf{26.0\% }& 5.94 $\rightarrow$ \textbf{7.62} & 26.89 $\rightarrow$ 28.30 & 0.63 $\rightarrow$ 0.64 &  66.87\% & 0.74 $\rightarrow$ 0.66 \\
\hline

DKITCHEN & 4.7\%$\rightarrow$7.0\% & 5.94 $\rightarrow$ 6.66 & 26.89 $\rightarrow$ 23.85 & 0.63 $\rightarrow$ 0.66 &  70.51\% & 0.74 $\rightarrow$ 0.71 \\
\hline

STRAFFIC & 4.7\%$\rightarrow$\textbf{30.3\%} & 5.94 $\rightarrow$ 7.40 & 26.89 $\rightarrow$ 31.77 & 0.63 $\rightarrow$ 0.61 &  64.68\% & 0.74 $\rightarrow$ 1.01 \\
\hline
\end{tabular}
} 
\end{table*}

\textbf{Environment Perturbations}: We present evaluations on three type of noises: indoor noise, background speaker noise, and outdoor noise, as they serve as representative examples of real-world acoustic environments. Unlike adversarial noise, environment perturbations primarily affect intelligibility, MCD, and timbre similarity in voice conversion output. As shown in Table \ref{tab:Environment Perturbations Evaluation}, intelligibility experiences a significant decline, with the highest Word Error Rate (WER) increase for STRAFFIC, rising from 4.7\% to 30.3\%, and the smallest increase for DKITCHEN. Among the three noise conditions, MCD values notably increase, with the smallest change for DKITCHEN (from 5.94 to 6.66) and the largest for STRAFFIC (from 5.94 to 7.40). The cosine similarity between the output and original utterance drops below 70\%, indicating a substantial difference between speech generated in noisy environments and the original.  However, Fundamental frequency-based metrics and subjective perception maintain a stable initial value. In summary, outdoor noise has the most disruptive effect on the Voice Conversion model, followed by background speaker noise, and then indoor noise.

\begin{table*}[!htbp]
\centering
\caption{Evaluation On Reverberant Conditions}
\label{tab:Reverberant Conditions}
\resizebox{\textwidth}{!}{ 
\begin{tabular}{|c|c|ccc|c|c|} 
\hline
\multirow{2}{*}{\textbf{Reverberant Conditons}} & \textbf{Intelligibility} & \multicolumn{3}{c|}{\textbf{Naturalness}} & \textbf{Timbre Similarity} & \textbf{Subjective Perception} \\
\cline{2-7} 
& \textbf{WER} & \textbf{MCD} & \textbf{F0RMSE} & \textbf{F0CORR} & \textbf{SS} & \textbf{UTMOS} \\
\hline
Room Size 5$\times$4 $\times$6 & 4.7\%$\rightarrow$\textbf{12.9\%} & 5.94 $\rightarrow$ 5.64 & 26.89 $\rightarrow$ 34.94 & 0.63 $\rightarrow$ 0.55 &  74.83\% & 0.74 $\rightarrow$ \textbf{0.46} \\
\hline

\end{tabular}
} 
\end{table*}

\textbf{Reverberant Conditions}: We present evaluations on reverberant audio simulated in a room size of 5$\times$6$\times$8. Similar to environmental perturbations, reverberant conditions introduce channel transmission interference that primarily impacts intelligibility and timbre similarity. As shown in Table \ref{tab:Reverberant Conditions}, the Word Error Rate (WER) moderately increases from 4.7\% to 12.9\%, and the cosine similarity between the output and original utterance is 74.83\%, indicating a difference between the utterance generated in reverberant conditions and the original utterance. Additionally, UTMOS score decrease by approximately 0.3, suggesting that the audio generated in reverberant conditions substantially affects the subjective perception.


\subsection{Case Study}
After evaluating the three types of input manipulation scenarios, we  seek to gain a deeper understanding of their specific effects on the intelligibility converted utterance. To this end, we select samples exhibiting substantial changes in Word Error Rate (WER) the total of 250 audio pairs, as detailed in Table \ref{tab:CaseStudy}.

Generally, there are fewer examples with significant WER changes before and after adversarial attacks and reverberant condition, while more such examples are found in the environment perturbation scenario. The key takeaway from the adversarial attack and reverberant condition scenarios is that interference with the input audio can lead to transcription errors in the output, such as spelling mistakes or the substitution of phonetically similar words. For example, in the adversarial attack scenario, the phrase "at home" is transcribed as "abnormal," "will" as "well," and "work" as "world." Similarly, in the reverberant condition, the phrase "his disappointment" is transcribed as "the boy."

However, unlike the first two types of input manipulation, environmental perturbations can lead to significant, sentence-level transcription errors in the output audio. For example, as shown in Table \ref{tab:CaseStudy}, the original phrase "the BBC was a disgrace" is completely altered under all noise conditions, including indoor, outdoor, and background speaker noise.  It can be observed that the output audio significantly diverges from the original meaning, with one example transcribed as "maybe it wasn't a surprise.". Notably, indoor and background speaker noise produce similar deviations, while outdoor noise results in a more pronounced change, transforming the phrase into "and that means hey it was a nice guy."

\begin{table*}[!htbp] 
\centering
\caption{Comparison of Original and Perturbed Utterance Transcription Output Under Three Types Of VC Input Manipulation}
\label{tab:CaseStudy}
\resizebox{\textwidth}{!}{ 
\begin{tabular}{ccccc}
\toprule
\textbf{Scene} & \textbf{WER Change} & \textbf{Content Text} & \textbf{Original Converted Transcript} & \textbf{Perturbed Converted Transcript} \\
\midrule
Adversary & 0.0\%$\rightarrow$57.14\% & but it is different in different regions & but it is different in different regions & \textcolor{red}{washington is deaf and} different region \\
Adversary & 0.0\%$\rightarrow$28.57\% & you can feel at home in china & you can feel at home in china & you can feel \textcolor{red}{abnormal} in china \\
Adversary & 0.0\%$\rightarrow$28.57\% & it will affect their work & it will affect their work & \textcolor{red}{did it well decide} their \textcolor{red}{world} \\
STRAFFIC & 0.0\%$\rightarrow$100.0\% & the bbc was a disgrace & the bbc was a disgrace & \textcolor{red}{and that means hey it was a nice guy} \\
OMEETING & 0.0\%$\rightarrow$80.0\% & the bbc was a disgrace & the bbc was a disgrace & \textcolor{red}{maybe it wasn't} a \textcolor{red}{surprise} \\
DKITCHEN & 0.0\%$\rightarrow$80.0\% & the bbc was a disgrace & the bbc was a disgrace & \textcolor{red}{then minb wasn't} a \textcolor{red}{surprise} \\
Reverberant & 0.0\%$\rightarrow$33.3\% & his appointment was generally welcomed yesterday & his appointment was generally welcomed yesterday & \textcolor{red}{the boy} was generally welcomed yesterday \\

\bottomrule
\end{tabular}
} 
\end{table*}

\section{Challenges and Outlook}\label{sec:open_issue}
As discussed above, we still have a long way to reach a robust VC system. Hence,  it is important to outline potential areas and directions for future research. Below are some key considerations for further exploration.


 \begin{itemize}
     \item \textbf{More explorations on adversarial noises generation}: Existing works \cite{huang_defending_2021,li_voice_2023} have explored  adversarial threat models in the frequency and time domains, primarily focusing on encoder-decoder based VC model architectures and applying perturbations only to the reference utterance. Therefore, measurements should be conducted to assess the effectiveness of attacks to GAN-based, VAE-based, Diffusion-based\cite{popov2021diffusion} or even Codec-based\cite{baade2024neural} VC models. In addition, attempts can also be made to apply perturbations on other input speech components in particular tasks, such as source utterance, emotion information, and accent changes.
     \item \textbf{More realistic audios on natural perturbation designing}: Existing methods primarily simulate real-world natural perturbations by mixing additive noise or reverberation noise with clean VC datasets. It is worth considering  the robustness evaluations of current VC models under environmental perturbations or reverberant conditions scenarios directly on real degraded speech data such as CHiME and WHAMR!.
     \item \textbf{More consistent semantic constraint optimization for speech disturbance}:Existing defenses mainly optimize signal domain protection limited to the semantic level \cite{2024SongBsAb}. Future work can be further expanded to study controllable perturbations at the levels of multidimensional paralanguage information, style features, and music structure, thereby achieving a comprehensive multimodal security protection system that spans semantics, melodies, and singing forms, covering the entire pre - and post generation scene. To ensure that the text content and semantic information remain stable even after incorporating defensive perturbations.
     \item \textbf{Better cascade defense methods}: Current passive defense strategies against noise interference mainly eliminate noise before feeding into the VC model by cascading state-of-the-art models, such as DEMUCS \cite{Défossez_Synnaeve_Adi_2020}, MetricGAN+ \cite{Fu_Yu_Hsieh_Plantinga_Ravanelli_Lu_Tsao_2021}, and Conv-TasNet \cite{Luo_Mesgarani_2019}. However, top models for speech enhancement or speech separation tasks inevitably cause some loss of feature information such as pitch, formant, energy, rhythm, and timbre while pursuing speech quality. It, to some extent,  distorts the converted speech generated by VC model. In addition, cascading models leads to an increase in the number of parameters, making it less suitable for deployment on mobile devices or real-time inference. Therefore, addressing the distortion of the cascaded models and managing the increase in model parameters is a worthwhile research topic for future passive defense solutions.
     \item \textbf{Better decoupling features in proactive defense methods}: Existing proactive defense strategies primarily focus on learning noise-invariant representations through denoising training or adversarial training to adapt to degraded speech data. However, these methods come with high training costs, offer defense limited to specific attacks, and pose risks of speech information leakage. Therefore, it is necessary to improve the decoupling ability of features such as pitch, content, and prosody.
     \item \textbf{More diverse evaluation metrics}: Existing attack strategies \cite{liu2023protecting,yu2023antifake,chen2024proactive} have measured imperceptibility and transferability on classic VC models. However, these indicators do not provide quantitative expressions, so it is necessary to objectively quantify these concepts. In addition, different data sources evaluation is also essential to confirm the effectiveness of attack strategies.
     \item \textbf{Focusing more on target oriented preference alignment optimization features}:Existing strategies focus on improving data and models to ensure that perturbations are applied to attacks on the model, and there is information inaccuracy involved in adversarial attacks against specific targets \cite{2025Advancing}. In future work, the potential distribution space of speech can be re modeled after adding noise, so that the perturbed space can still converge to the optimal region of human true speech intelligibility. By establishing primary controllable priors through SFT and using RL algorithm for preference alignment, human subjective preference feedback is explicitly constrained to the noise distribution mapping process, achieving robust consistency and generalization diversity in multiple domains.
     \item \textbf{The defense mechanism of watermark embedded anti-counterfeiting that focuses more on noise}:The existing main adversarial attacks are basically introduced by embedding identifiable steganographic markers in the spectral features through watermark embedding, to verify the authenticity and integrity of the speech source \cite{2023WavMark}. In the future, this mechanism can detect adversarial sample attacks without significantly affecting speech quality, and then process them through adversarial training, encoder denoising and reconstruction methods. At the same time, it can maintain strong robustness detection ability even after being affected by noise interference or model tampering, effectively improving the passive security defense level of VC systems in open environments.
     \item \textbf{Prompt strategies in robust VC systems}: Recently, the latest work in speech synthesis \cite{yao2023promptvc,guo2023prompttts,leng2023prompttts,shimizu2023prompttts++,xin2024ralle} attempts to generate style vectors driven by natural language prompts as conditions. It improves control ability as well as enhances the interpretability of the model. Similarly, robust VC system can be guided by natural language, enabling the model to learn noise-resistant representations. It helps improve the adaptability of VC systems to the degraded speech data after semantic-level and acoustic-level perturbation.
     \item \textbf{Exploring the connection between speech language models and VC systems}: Recent work  has explored the integration of VC with other tasks. SpeechComposer \cite{wu2024speechcomposer} utilizes a fixed set of tokens to enable knowledge sharing between tasks. Make-A-Voice \cite{huang2023makeavoice} has established a unified framework for three synthesis tasks: TTS, VC, and singing voice synthesis (SVS), which allows for synthesize and manipulate voice signals from discrete representations. The S2ST framework \cite{wang2023speech} draws on the idea of zero-shot style transfer in VC to successfully preserve the speaker's timbre in voice translation tasks. Thus, it highlights the growing need to explore the connections between VC and other speech tasks.
 \end{itemize}


\section{Conclusions}\label{sec:conclusions}
In this paper, we present a survey of attack and defense strategies for voice conversion (VC). We categorize various robust VC papers from the past five years according to adversarial attacks, environmental perturbations, and reverberation conditions, and organize corresponding proactive and passive defense strategies. In addition, this survey covers various clean and noisy datasets and organizes evaluation metrics for quantifying robust VC systems from the perspectives of naturalness, intelligibility, timbre similarity and subjective perceptions. Finally, considering the limitations and gaps in existing VC system attack and defense strategies, we propose some suggestions for future work. Therefore, this survey provides a solid foundation and motivation for future research on developing robust and secure neural networks for VC tasks.

\begin{acks}
The work is partially supported by the National Nature Science Foundation of China (No. 62376199, 62206170, 62376246). Besides, we express our gratitude to the authors of AttackVC and VoiceGuard for providing the source code for our experiments.
\end{acks}

\bibliographystyle{ACM-Reference-Format}
\bibliography{sample-base}

\appendix









\end{document}